\documentclass[preprint,review,12pt]{elsarticle}
\usepackage{graphicx}
\usepackage{epsf}
\usepackage{amsmath}
\usepackage{lscape}
\usepackage{color}
\usepackage{dcolumn}
\usepackage{hyperref}
\usepackage{amsmath}
\usepackage{rotating}
\usepackage{soul}
\usepackage{longtable}
\usepackage{natbib}

\newcommand{\cm}{cm$^{-1}$}

\newcommand{\etal}{\emph{et al}.}

\def\Xstate{{X~$^{1}\Sigma_{\rm g}^{+}$}}
\def\Astate{{A~$^{1}\Pi_{\rm u}$}}

\def\Bstate{{B~$^{1}\Delta_{\rm g}$}}

\def\astate{{a~$^{3}\Pi_{\rm u}$}}
\def\bstate{{b~$^{3}\Sigma_{\rm g}^{-}$}}
\def\cstate{{c~$^{3}\Sigma_{\rm u}^{+}$}}
\def\dstate{{d~$^{3}\Pi_{\rm g}$}}

\newcommand{\lande}{Land\'e}

\newcommand{\ket}[1]{\vert #1 \rangle  }

\journal{Journal of Molecular Spectroscopy}

\begin{document}
	
	\begin{frontmatter}
		
		\title{Predicted Land\'e $g$-factors for open shell diatomic molecules}
		\author{Mikhail Semenov, Sergei.~N. Yurchenko, Jonathan Tennyson}
		\address{Department of Physics and Astronomy, University College London,
			London, WC1E 6BT, UK}
		
		\begin{abstract}
The program {\sc Duo} (Yurchenko {\it et al.}, Computer Phys. Comms., 202
(2016) 262) provides direct solutions of the nuclear motion Schr\"odinger
equation for the (coupled) potential energy curves of open shell
diatomic molecules. Wavefunctions from  {\sc Duo} are used to compute
Land\'e $g$-factors valid for weak magnetic fields; the results
are compared with the idealized predictions of both Hund's case (a)
and Hund's case (b) coupling schemes. Test calculations are performed
for AlO, NO, CrH and C$_2$. The computed $g_J$'s both provide a sensitive test
of the underlying spectroscopic model used to represent the
system and an indication of whether states of the molecule are
well-represented by the either of the  Hund's cases considered.
The computation of Land\'e $g$-factors is implemented as a standard
option in the latest release of  {\sc Duo}.			
		\end{abstract}
		
	\end{frontmatter}
	
	\section{Introduction}
	
The lifting of the degeneracy of the energy levels in molecule by a magnetic field is
a well-known and well-studied phenomenon. Thus it has spawned experimental
techniques such as laser magnetic resonance spectroscopy \cite{81Davies,06GoRoBe.CoH},
magnetic rotation spectroscopy \cite{92MccFie.NiH},
and optical Zeeman spectroscopy \cite{14ZhStxx.MgH,12QiLiSt.TiH}. These techniques,
for example, use the Zeeman effect to tune transitions in and out resonance
by changing the applied magnetic field \cite{07ChBaPe.CrH}.
Zeeman effects can also be probed directly using standard spectroscopic
techniques to study molecules in magnetic fields \cite{09VaAsCr.NiH,12RoCrRi.NiH,14CrDoRi.FeH}.
In a similar
fashion, Zeeman effects are increasingly being used to form, probe and trap
molecules at ultra-cold temperatures  \cite{98WeDeGu.CaH} for example by
use of  magnetically tunable Feshbach resonances \cite{06KoGoJu}.

Spectral shifts and splittings provide a remote sensing technique with
which to study the Universe. Zeeman splitting of molecular spectra are
actively being used to probe magnetic fields in a variety of
astronomical environments including sunspots
\cite{00BeFrSo.TiO,01BeSoxx.OH,Sunspot_measurement_problem}, starspots
\cite{15AfBexx}, white dwarfs \cite{Zeeman_In_White_Dwarfs,15FeDeGa}
M-dwarfs \cite{10ShReWe.FeH} and potentially exoplanets
\cite{Zeeman_exoplanets_velocity}.

The Zeeman splitting patterns of the spectrum of an open shell diatomic
molecule can be calculated in a straightforward fashion provided
that the quantum numbers characterizing states in question are known
and are conserved. However, there are circumstances, such as resonance
interactions between nearby states via spin-orbit or other couplings where the quantum numbers used to specify the
electronic state
associated with a given level are not precisely conserved. In this
case evaluation of Zeeman splitting as represented by the Land\'e $g$-factor
is not straightforward and requires a numerical treatment. It is such a
treatment which is the focus of the present article.

The important advantage of the Zeeman effect is that the associated
splitting can be made large enough to separate otherwise degenerate
spin-components ($\Lambda$-doublet). Moreover, the measurement of the $g$
values can be more accurate than the energy spacing
\citep{90GrLiFi.NiH}.

The Zeeman methodology for diatomics was introduced by Schade
\cite{78Schade.methods}.  Stolyarov {\it et al.}  \cite{92StKlTa.Na2}
investigated the perturbations in the calculation of the \lande\
factors caused by interactions with other electronic states. In a very recent theoretical work Borkov {\it et al.}
\cite{16BoSuKl.NO}
presented a numerical model of Zeeman splitting based on the
use of effective molecular Hamiltonians.

%From discussion by \cite{90GrLiFi.NiH}: ``The molecular Zeeman effect has advantageous diagnostic
%power in that it can usually be made large enough to
%be resolved with a cw laser-based experimental scheme
%whereas A doubling is often too small for splittings in low-J
%lines to be resolved. Furthermore, the experimental determination
%of g values from Zeeman spectra can be somewhat
%more accurate than measuring individual A-doublet spacings
%simply because of the redundancy of measuring several
%equivalent magnetic sublevel spacings for each rotational
%level.''

Le Roy's {\sc LEVEL} \cite{level} has become the program of choice for
solving the diatomic nuclear motion problem. However, {\sc LEVEL} can
only treat open shell molecules in limited circumstances
\cite{15WaSeLe.NaH,jt605} and does not consider the coupling between
states by spin-orbit and related effects which can have an important
effect of the $g$-factors.  For this reason we have written our own
diatomic nuclear motion code {\sc Duo} \cite{jt609}. {\sc Duo}
explicitly treats open shell systems and can allow for coupling
between the various states involved.  In this context, of particular
interest to us are the many open-shell diatomic molecules which are
known to be present, or may be present, in hot astronomical atmospheres
such as those found in cool stars and exoplanets.
Such species are being studied  as part of the ExoMol
project \cite{jt528}. Zeeman splittings in these molecules can provide
useful information on the magnetic fields present in these distant
bodies. So far ExoMol has created spectroscopic models for a number of
open shell diatomic species \cite{jt605,jt598,jt599,jt618,jt644,jtCrH,jtTiH,jtPS,jtTiO}.

In this paper we present extension to  {\sc Duo} which allows
Land\'e $g$-factors to be computed for individual states of open shell
systems. As initial examples we focus on
four systems studied by ExoMol, namely
AlO  \cite{jt598}, NO \cite{jtNO}, CrH \cite{jtCrH} and C$_2$ \cite{jt637,jtexoC2}.
These systems were selected as ones of interest for ExoMol and
for which there are laboratory Zeeman spectra. These laboratory
studies are discussed below.

\section{Theory}

Discussions of the underlying theory and methodology used in {\sc Duo}
is given elsewhere \cite{jt609,jt589,jt632} so only key points are considered
below. {\sc Duo} solves the diatomic nuclear motion problem using
a Hund's case (a) basis. This does not represent
an approximation even for molecules poorly represented
by  Hund's case (a) since  a complete set of angular momentum functions
are used for a given total angular momentum, $J$. We note that the same
choice has been adopted by others \cite{95Marian.NiH,15Schwenke.diatom}.

The basis set used by {\sc Duo} can be written as
\begin{equation}\label{e:basis}
 \ket{n} =  \ket{{\rm state}, J, \Omega, \Lambda, S, \Sigma, v }  =  \ket{ {\rm state},
\Lambda, S, \Sigma  } \ket{ J,\Omega,M  \rangle |{\rm state}, v },
\end{equation}
where  $\Lambda$ is the projection of electron angular momentum on
the molecular axis;
$S$ is the electron spin quantum number with projection
$\Sigma$ along the  molecular axis and  $\Omega$ is corresponding
projection of $J$. The vibrational quantum number is given
by $v$ and the label `state' is used to  denote
the electronic state which is required for both the state-dependent
angualar momenta and the vibrational state.
$M$ is the projection of the total angular momentum along the
laboratory axis $Z$ and is therefore the magnetic quantum number
which quantizes the splitting of the levels in a weak
magnetic field. Finally, $\ket{n}$ is simply a compound index representing
the various quantum numbers. These basis functions are symmetrized
to give a definite parity, $\tau$. Only $J$ and $\tau$ are conserved
quantum numbers with the addition of  u/g for homonuclear molecules.

{\sc Duo} obtains the wavefunctions for a given nuclear motion problem
by diagonalizing a coupled-states Hamiltonian. These wavefunctions,
$\phi_{\lambda}^{J\tau}$, are then given by
\begin{equation}
\label{equation:DUO_wavefunction}
\phi_{\lambda}^{J\tau}=\sum_{n}{C_{\lambda n}^{J\tau}\ket{n}},
\end{equation}
where $\lambda$ denotes the electronic state.

In the case of weak magnetic fields, the Zeeman splitting can be approximated by
\begin{equation}\label{e:Zeeman}
  \Delta E_{\rm B} = g_J M \mu_0 B,
\end{equation}
where $\Delta E_{\rm B}$ is the shift in energy of a state with total
angular momentum $J$ and projection of $J$ along the field direction
is $M$,  $g_J$ is the \lande\ factor, $\mu_0$ is the Bohr magneton,
$B$ is the magnetic field.  Within a Hund's case (a) representation,
the Land\'e $g$-factor is given by
\cite{90GrLiFi.NiH,02BeSoxx.diatom}:
\begin{equation}
\label{equation:lande_hunds_a}
g_J^{\rm (a)}=\frac{(g_L\Lambda+g_S\Sigma)\Omega }{J(J+1)}
\end{equation}
where $g_S$ and $g_L$ are the standard
electron spin and orbital $g$-factors respectively. If $\Lambda$ and
$\Sigma$ are conserved quantities for a given rovibronic state then
this expression is analytic; below this will be known as the
QN(a) approximation meaning good  quantum numbers in Hund's case (a).

The good-quantum number approximation has been also introduced and
used in the case of the NiH spectroscopy by Gray {\it et al}
\cite{90GrLiFi.NiH}. Here we used $g_J$ as a total \lande\ factor
which includes all other contributions from the electron spin and
orbital angular momenta; this is different from the definition
conventionally used in Zeeman experimental studies, see, for example,
Gray {\it et al} \cite{90GrLiFi.NiH}.

The corresponding expression for the Hund's case (b) \lande\ factor
can be approximated by \citet{02BeSoxx.diatom}
\begin{eqnarray}
\label{equation:lande_hunds_b}
g_J^{\rm (b)} &=& \frac{g_L}{2 J(J+1)} \left\{ \frac{\Lambda^2 \left[ J(J+1) + N(N+1) -S(S+1) \right] }{N (N+1)}
\right\} \notag \\
&+&  \frac{g_S}{2 J(J+1)}\left[ J(J+1) - N(N+1) +S(S+1)\right]  ,
\end{eqnarray}
%\red{can someone check this.}
where $N$ is the rotational quantum number. If $\Lambda$ and
$N$ are conserved quantities for a given rovibronic state then
this expression is analytic; below this will be known as the
QN(b) approximation meaning good  quantum numbers in Hund's case (b).

The intermediate (and more general) case can be modeled using the
$\underline{G}$ matrix with the following matrix elements
\cite{02BeSoxx.diatom}:
\begin{eqnarray}
% \nonumber % Remove numbering (before each equation)
  G_{\Sigma,\Sigma} &=& \frac{(g_L\Lambda+g_S\Sigma)\Omega }{J(J+1)}, \\
  G_{\Sigma,\Sigma\pm 1} &=& g_S \frac{\sqrt{S(S+1) - \Sigma (\Sigma\pm 1)} \sqrt{J(J+1) - \Omega (\Omega\pm 1)}
  }{2 J(J+1)} \delta_{v,v'} \delta_{\Lambda,\Lambda'} \delta_{S,S'}.
\end{eqnarray}

In practice $\Lambda$ and $\Sigma$ are not generally conserved when spin-orbit and other
curve coupling effects are taken into account. In this case one can
use the {\sc Duo} wavefunctions to compute $g_J$ for a given rovibronic
state by averaging over the corresponding wavefunction as given by
\begin{equation}
\label{equation:lande_hunds_diag}
G_{\Sigma,\Sigma}^{\rm Duo} = \sum_n  |C_{\lambda n}^{J\tau}|^2 \frac{(g_L\Lambda_n+g_S\Sigma_n)\Omega_n }{J(J+1)},
\end{equation}
\begin{eqnarray}
\label{equation:lande_hunds_offdg}
\nonumber
  G_{\Sigma,\Sigma\pm 1}^{\rm Duo} =\sum_n \sum_{n'} {C_{\lambda n}^{J\tau}}^{*} C_{\lambda n'}^{J\tau}
  \delta_{v,v'} \delta_{\Lambda,\Lambda'} \delta_{S,S'} && \\
\times g_S \frac{\sqrt{S_n(S_n+1) - \Sigma_n (\Sigma_n\pm 1)} \sqrt{J(J+1) - \Omega_n (\Omega_n\pm 1)}  }{2 J(J+1)}
,
\end{eqnarray}
where $S_n$, $\Lambda_n$ and $\Sigma_n$ are the values of $S$, $\Lambda$ and $\Sigma$
taken in basis function $\ket{n}$ and  $\Omega_n = \Lambda_n + \Sigma_n$.
%This approach also relies on the approximation of the large separation of the field-free energies within Hund's
case (a).
In the following we also assume  $g_L=1$ and $g_S=2.0023$.
%\red{did we use the correct value for $g_S$ and not just 2?}

To help interpret our results we
define the difference between $g_J$'s evaluated using
{\sc Duo} wavefunctions and using the QN approximation as given by.
\begin{equation}\label{e:Lande-diff}
 \Delta g_J^{\rm (x)} =  g_J^{\rm Duo} - g_J^{\rm (x)}.
\end{equation}
where (x) is (a) or (b) as appropriate.

\section{Results}

\subsection{AlO}

As an initial test case Land\'e $g$-factors were computed for aluminium
oxide (AlO). The spectroscopic model used for AlO is due to Patrascu
{\it et al.} \cite{jt598,jt589} which is based on {\it ab initio}
curves tuned to reproduce the extensive set of experimental spectra.
The model comprises three electronic states: X~$^2\Sigma^+$,
A~$^2\Pi$ and  B~$^2\Sigma^+$. The latter two states lie, respectively,
5406 and 20688 \cm\
above the ground states. The closeness of the X and A states leads to
significant mixing. A recent study of radiative lifetime \cite{jt624}
using this model showed strong effects due to X -- A mixing but that the
B state appeared largely unperturbed.

\begin{figure}[htbp]
		\centering
\scalebox{0.6}{\includegraphics{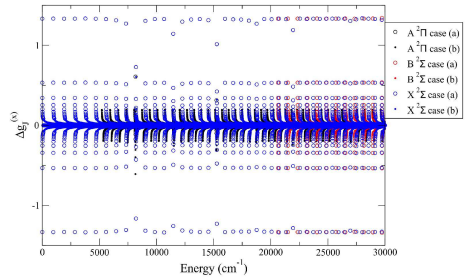}}
\caption{Difference between Land\'e $g$-factors obtained for AlO using {\sc Duo} wavefunctions and the QN
approximations.}
\label{fig:AlO}
\end{figure}

Figure~\ref{fig:AlO} shows the difference between $g_J$'s evaluated
using {\sc Duo} wavefunctions and using the QN approximation These
results show systematic effects. Firstly, it would appear that the
$g_J$-factors for
the   X~$^2\Sigma^+$ and  B~$^2\Sigma^+$ are significantly better
represented in the Hund's case (b) than case (a). Secondly, the
A~$^2\Pi$ appears closer to  Hund's case (a), although in this case
the differences are smaller. Finally,
there are a number
perturbations caused by a coupling between levels in the X and A states.
Such interactions have been noted before \cite{jt624}.

\begin{figure}[htbp]
		\centering
\scalebox{0.4}{\includegraphics{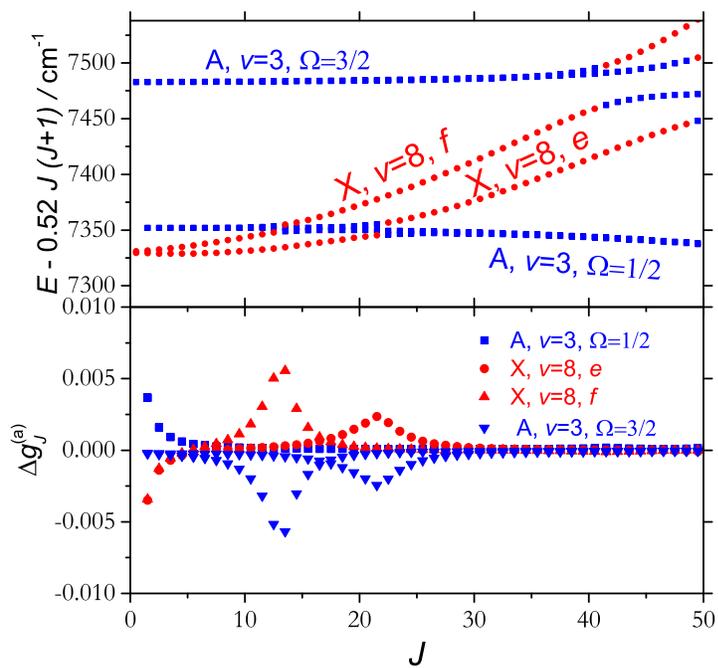}}
\caption{Reduced energy term values of AlO in the region of  the crossing between X, $v=8$ and A, $v=3$ (upper
display)
and difference between Land\'e $g$-factors obtained for AlO using {\sc Duo} wavefunctions and the QN Hund's case
(a) approximation.}
\label{f:coupling}
\end{figure}

\begin{figure}[htbp]
		\centering
\scalebox{0.4}{\includegraphics{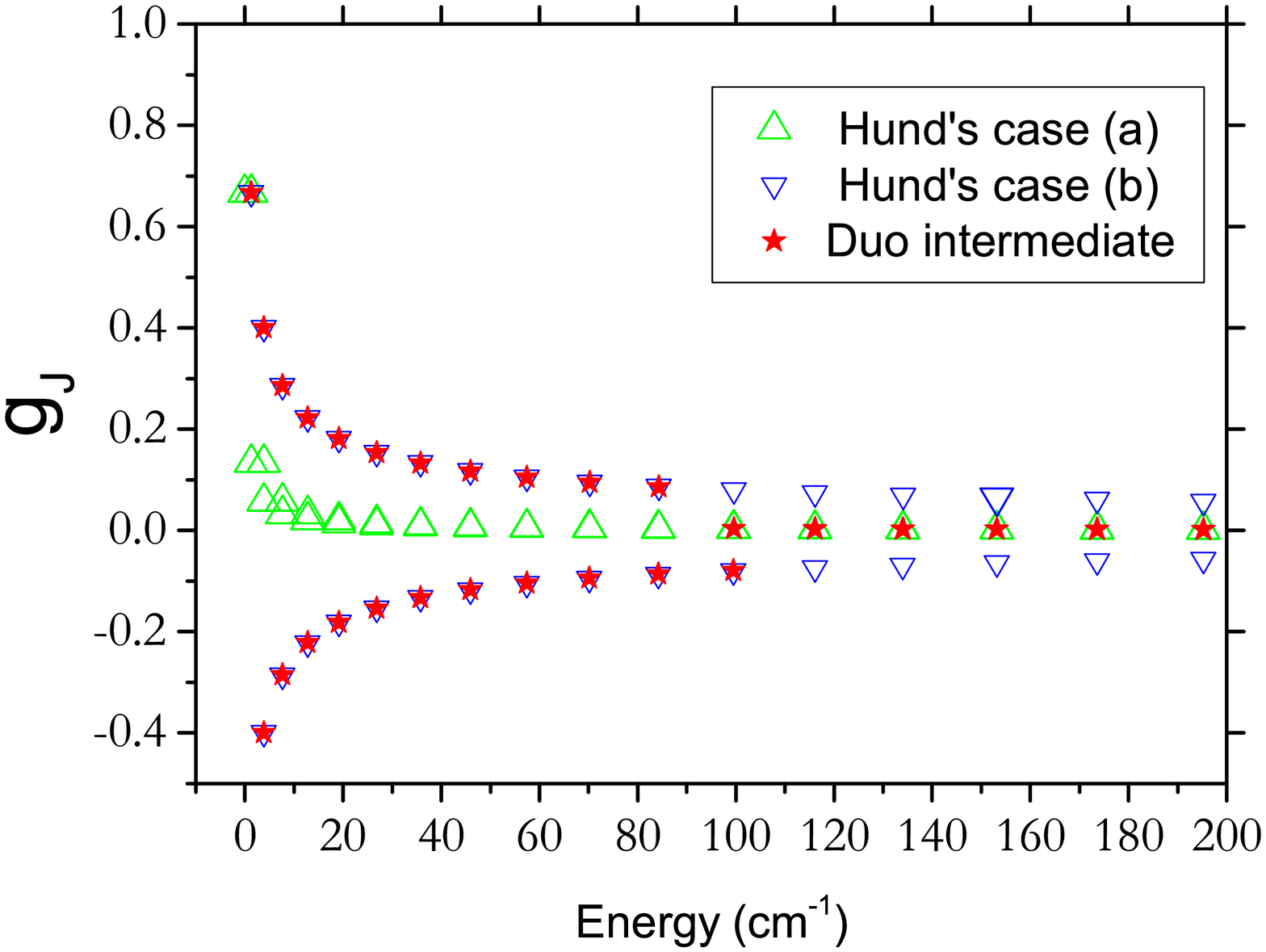}}
\caption{Land\'e $g$-factors of AlO obtained using  {\sc Duo} wavefunctions with
Eqs.~(\ref{equation:lande_hunds_diag},\ref{equation:lande_hunds_offdg}) and the QN approximation, Hund's cases (a)
and (b). }
\label{f:AlO:Duo:a:b}
\end{figure}

Changes due to  the X -- A state coupling
appearing as well pronounced structures in Fig.~\ref{fig:AlO} suggest
that the B state $g$-factors are largely unchanged from the idealised
values. In order to illustrate how the coupling between different
states affect the values of the \lande\ factors in
Fig.~\ref{f:coupling} we show energy crossings between the X ($v=8$)
and A ($v=3$, $\Sigma = 1/2$ and $\Sigma = 3/2$) rovibronic states
(upper display). The spikes in the $J$ progressions of the \lande\
factors (lower display) appear at the same $J$ values (13.5 and 21.5)
as the two crossings.  This is where the wavefuncitons are extremely
mixed and the quantum number approximation,
Eq.~(\ref{equation:lande_hunds_a}), becomes very poor.
The net effect   from the {\sc Duo} model even in case of strongest couplings of AlO
is still relatively small, of the order of $10^{-2}$ for $J=50$
see Fig.~\ref{f:AlO:Duo:a:b}.

Figure~\ref{f:AlO:Duo:a:b} shows the $g_J$ values computed using the
three methods: Duo wavefunctions, Hund's case (a) and Hund's case (b)
approximations for the X ($v=0$) state. The character of $g_J$ changes
from (b) to (a) as energy increases, illustrating importance of the
proper modelling of the Zeeman effect.

We should note the study by Gilka {\it et al.} \cite{08GiTaMa.AlO}, where the effects of couplings between orbital-
and spin-angular momenta of the X and A states on the $g_S$ values were also studied in their {\it ab initio}
calculations of the $g$-tensor of AlO.

\subsection{NO}

The nitric oxide (NO) molecule provides a rather simple test of our
methodology. McConkey  {\it et al.} \cite{jtNO} recently constructed
a spectroscopic model and generated the associated line list for NO
considering only the X~$^2\Pi$ electronic ground state. McConkey  {\it et al.}
consider all 6 major isotopologues of the system; here we restrict
ourselves to $\mathrm{^{14}N^{16}O}$ for which
there are some limited,
experimental studies on the behavior of its ground state ro-vibrational
transitions  in a magnetic field,
albeit a relatively strong one \cite{11IoKlKo.NO}. These observations
have been subject of recent models \cite{16BoSuKl.NO} which show
that for most of the field strengths considered it was necessary
to move beyond the linear Zeeman effect considered here. We note the
Zeeman effect in NO has also been used to ascertain the
distribution of NO in the Martian atmosphere \cite{08CoSaGe.NO}.

The spectroscopic model of  McConkey  {\it et al.} \cite{jtNO} made
extensive use of experimental data in determining both the shape of
the X~$^2\Pi$ potential energy curve and the various coupling terms.
However even with a high quality fit their model does
not predict transition frequencies with the accuracy required for
studying the relatively small Zeeman splittings. We therefore follow
McConkey  {\it et al.}  and adopt as our zero-field energy levels
the empirical values they determine.

The experimental study of Ionin {\it et al.} \cite{11IoKlKo.NO} only
considered in detail the splitting of the $^2\Pi_{3/2}$ Q(2.5)
fundamental transition at 1875.7228 cm$^{-1}$ as a function of
magnetic field. There strategy was to use the magnetic field to tune
the transition into resonance with CO laser lines.  Ionin {\it et al.}
observed 3 components of this transition, namely the $(M',M")$ ones
associated with $({1}/{2},-{1}/{2})$, $({3}/{2},{1}/{2})$
and  $({5}/{2},{3}/{2})$ which they observed using the CO
laser line $9 \rightarrow 8$ P(15) which lies at 1876.30  cm$^{-1}$.
Bringing the lines into resonance required magnetic fields of
approximately 3.8, 4.2 and 5.6 T respectively. Strong fields
are required since the $g$ factors for the ground and first excited
vibrational states are fairly similar so the transition frequency
only depends weakly on $B$.

If the Zeeman splittings were linear then the three lines considered
would all lie at the same frequency.  In practice this is only true up
to about 2 T  and for fields above this value
the quadratic Zeeman effects become increasingly important
\cite{11IoKlKo.NO,16BoSuKl.NO}. The previous studies suggest that
 only the $(M',M") =
({1}/{2},-{1}/{2})$ transition frequency varies approximately
linearly with field strength in the 2 -- 6 T region.
Our calculations place this transition at 1876.29 cm$^{-1}$ for
a field of 3.8 T, in excellent agreement with the observations.
This values are obtained from our calculated $g_J$ factors
of 0.316625 and 0.316857 for the $^2\Pi_{3/2}$ $J=2.5$ state of $v=0$ and $v=1$ respectively.

\subsection{C$_2$}

The carbon dimer is a very well studied system 	\cite{jt637,92Martin} whose
spectrum is widely used for studying astronomical and terrestrial plasmas.
The many systems  of C$_2$ electronic bands are well-known to have many
perturbations due to couplings between states, something that should
be reflected in the Land\'e $g$-factors.

The ExoMol model for C$_2$ considers the eight lowest electronic states
of C$_2$: \Xstate, \Astate, \Bstate, B$'$~$^{1}\Sigma_{\rm g}^+$,
\astate, \bstate, \cstate, and \dstate.
Due to strong interactions between rovibronic states in this system especially at high rotational excitations, the
quantum numbers becomes meaningful and therefore very difficult to correlate between Hund's cases. Therefore we
restrict ourselves to states with $J \leq 35$ and energies up to 35,000 \cm.

\begin{figure}[htbp]
		\centering
\scalebox{0.6}{\includegraphics{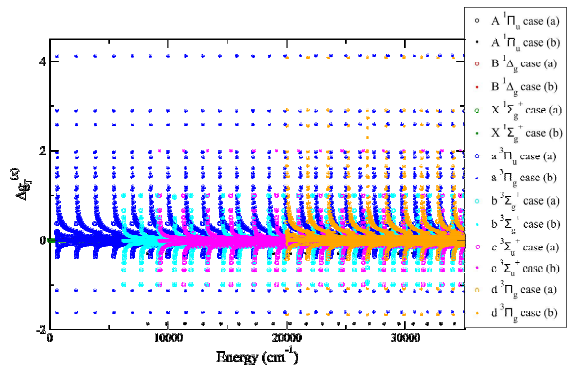}}
\caption{Difference between Land\'e $g$-factors obtained for C$_2$ using {\sc Duo} wavefunctions and the QN
approximation.}
\label{fig:C2}
\end{figure}

Figure \ref{fig:C2} gives an overview of our computed Land\'e $g$-factors for
C$_2$. It is clear that for this system the changes caused by coupling
between states are large.

% this is especially true for levels
%associated with the $1~^{5}\Pi_{\rm g}$
%state.

Apart from the characteristic spikes as in the case of AlO, these \lande\ factors show that these
deviated values also build well defined horizontal patterns. These patterns should indicate a deviation of
the C$_2$ spectra from Hund's case (a). The transition from Hund's cases (a) and (b) for different $J$ is
a well-known issue in the analysis of the rovibronic spectra of C$_2$, see, for example, \cite{94PrBe.C2}.

\section{CrH}

As a third example we consider the CrH molecule, another
astronomically important species. The ExoMol model \cite{jtCrH} for
this system considers the lowest 8 electronic states:
X~$^{6}\Sigma^+$, a~$^{4}\Sigma^+$, A~$^{6}\Sigma^+$, B~$^{6}\Pi$,
b~$^{4}\Pi$, C~$^{6}\Delta$, c~$^{4}\Delta$ and the lowest
dissociative $^8\Sigma^+$ state.  Here we consider states with $J
\leq 35$ and energies up to 35,000 \cm.
%We note that above 20,000 \cm\ the electronic states become increasingly dense.
As the model used does not consider electronic states
with thresholds above 20,000 \cm,
where new states become increasingly dense,
it is likely that our calculations will underestimate the perturbation of the $g$-factors in this region. However
even below 10,000 \cm, where the states all belong to X~$^{6}\Sigma^+$ electronic state the perturbations are
fairly large.

\begin{figure}[htbp]
		\centering
\scalebox{0.6}{\includegraphics{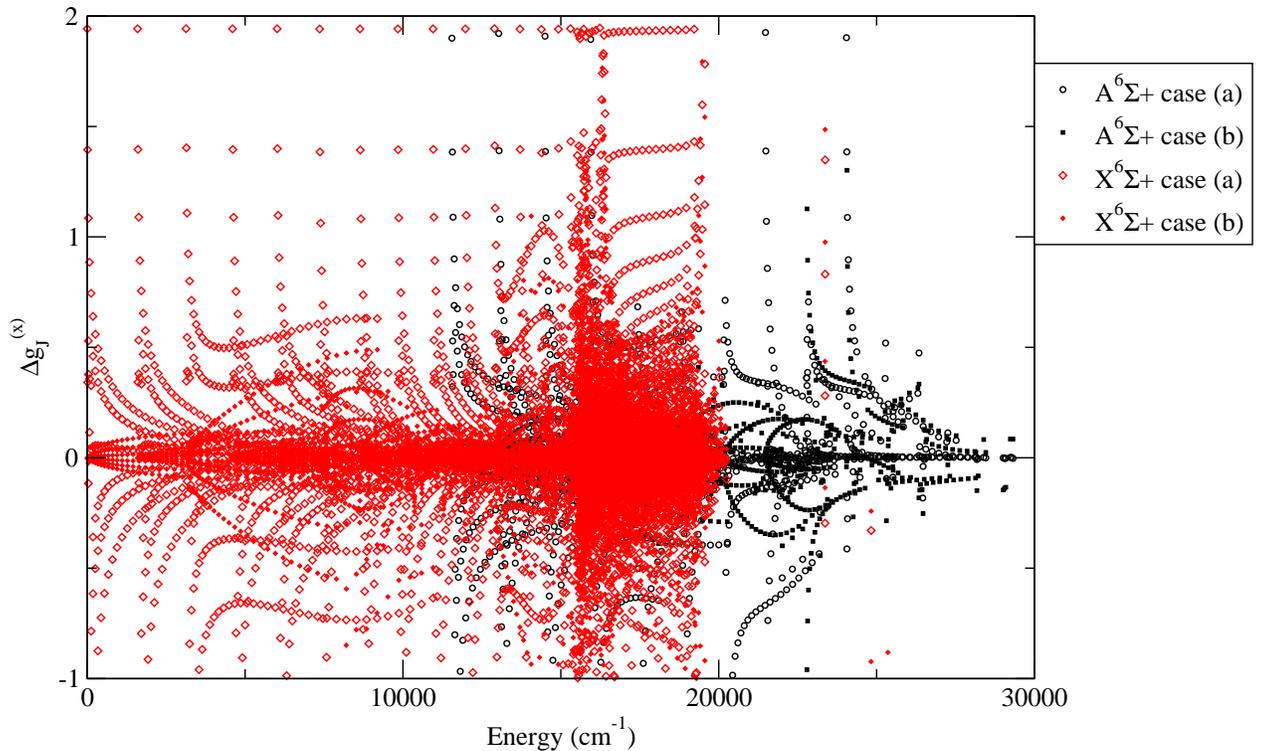}}
\caption{Difference between Land\'e $g$-factors obtained for CrH using {\sc Duo} wavefunctions and the QN
approximation.}
\label{fig:CrH}
\end{figure}

Figure \ref{fig:CrH} gives an overview of our computed Land\'e
$g$-factors for the lowest two sextet states
of CrH compared to the QN approximations. Clearly there
are lots of structures. For the X~$^{6}\Sigma^+$ state, QN~(a) appears a poor approximation
in all cases; QN~(b) does better but still shows pronounced structures
starting at about 4000 \cm. Above about 14000 \cm\ all X state $g_J$
factors appear highly perturbed. This is also true for the singlet A state.

Chen {\it et al.} \cite{07ChBaPe.CrH} measured effective $g$ values
for a few levels in CrH; Table~\ref{tab.CrH} compares these
measurements (A~$^{6}\Sigma^+$, $v=1$) with our results.
This table also shows that the Hund's  case (b) approximation to $g_J$ is more appropriate than
case (a). Chen {\it et al.}  \cite{07ChBaPe.CrH}  also reported an averaged $g$ value for
the A~$^{6}\Sigma^+$, $v=0$ state over the measured values for four
states $J=3/2, N=1$, $J=5/2, N=0$, $J=5/2, N=1$, $J=7/2, N=1$ of
  2.0081(20). Using the Duo $g_J$ values to produce the same averaging
  we obtained 1.9976, which is in very close agreement to the
  experiment. The agreement for other levels is not as good, suggesting
that our model for CrH needs further improvement.

\begin{figure}[htbp]
		\centering
\scalebox{0.4}{\includegraphics{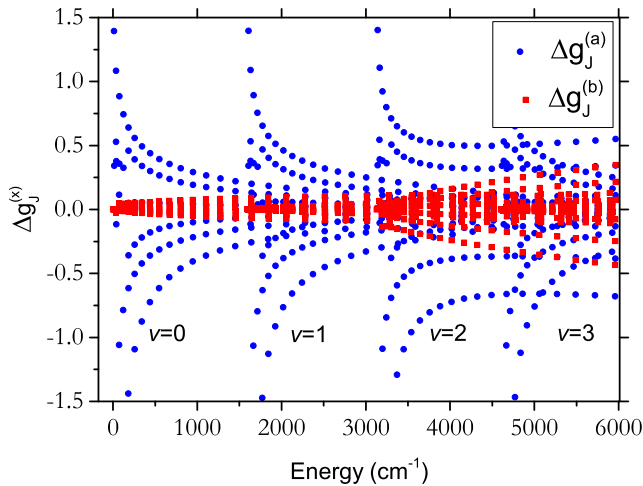}}
\caption{Difference between Land\'e $g$-factors obtained for CrH using {\sc Duo} wavefunctions, the QN case (a) and
(b) approximations.}
\label{f:CrH:A:B:Duo}
\end{figure}

\begin{table}
\caption{Comparison between measured values
of  $g_J$ for CrH due to Chen \etal\ \cite{07ChBaPe.CrH} and our calculated values using the Duo
wavefunction approach, Hund's cases (a) and (b) approximations. }
		\begin{center}
			\begin{tabular}{crrrl}
				\hline\hline
state &  \citep{07ChBaPe.CrH}  & Duo &  QN(a) & QN(b) \\
\hline
$N=0, J=5/2$	&	1.7468(17)	&	1.9781	&0.0571	&	2	\\
$N=1, J=3/2$	&	1.8760(30)	&	2.8168	&0.1333	&	2.8	\\
$N=1, J=5/2$	&	1.7208(28)	&	1.7433	&0.0571	&	2	\\
$N=1, J=7/2$	&	1.9123(25)	&	1.4221	&1.4286	&	1.7714	\\
\hline\hline
\end{tabular} \label{tab.CrH}
\end{center}
\end{table}

\section{Conclusion}

We have developed a numerical procedure for evaluating Land\'e $g$-factors
for diatomic molecules without making any assumptions about conserved
quantum numbers. This method is tested for four molecules AlO, NO, C$_2$ and
CrH. It would seem that besides making predictions for $g$-factors,
the comparison between our computed value and the value
predicted under the assumption of particular Hund's case gives a clear
means of distinguishing those levels which are best represented by
Hund's case (a) from those which are approximately Hund's case (b)-like.

The accuracy of our predicted $g_J$ factors depend on a number of factors:
(a) the accuracy of the underlying spectroscopic model used and, in particular,
its ability to reproduce coupling between different electronic states,
(b) our ability to solve this model by, for instance, converging
the basis set representation and (c) any assumptions made about
angular momentum couplings
within the system. Although our procedure is based on a Hund's case (a)
coupling scheme, our general formulation means that no actual approximations
are made by adopting a (complete) basis formulated within this scheme.
Similarly it is relatively easy, and computationally cheap, to use large
vibrational basis sets when converging the problem. This means that
the choice of spectroscopic model is likely to be the major source
of uncertainty in our calculations or, conversely, that available
measurements of Land\'e $g$-factors have the potential to be used
to improve the spectroscopic model. In addition we note that our
formulation is only appropriate when the changes depend linearly with
the magnetic field. Inclusion of non-linear effects require a more
sophisticated treatment which we plan to study in future in work.

It is our plan to routinely compute  Land\'e $g$-factors for all open
shell diatomic species studied as part of the ExoMol project from
now on. To this end, the new ExoMol data format \cite{jt631} has
been adjusted to include the computed values of $g_J$ for each state
as part of the states file made available for each isotopologue studied.

	\section*{Acknowledgements}
We thank Patrick Crozet for drawing our attention to mistakes in the published
version of this article. 	
This work was supported by the European Research Council under Advanced Investigator Project 267219 and the COST
action MOLIM (CM1405).
	
%\bibliographystyle{model1a-num-names}

%\bibliography{journals_phys,jtj,methods,linelists,diatomic,NiH,FeH,NaH,TiH,CaH,C2,MgH,TiO,Na2,NO,CoH,CrH,OH,MS,programs,AlO}

\end{document}